\documentclass[aps,prl,twocolumn,showpacs,a4paper,unsortedaddress,amsmath,amssymb,byrevtex]{revtex4}

\usepackage[latin1]{inputenc}
\usepackage{graphicx}
\usepackage{color}
\usepackage{dcolumn}
\usepackage{bm}
\usepackage{hyperref}
\usepackage{times}

\begin{document}

\title{Predictions for the formation of new atomic chains in Mechanically Controllable
Break Junction experiments}

\author{Lucas Fern\'andez-Seivane$^1$}
\author{V\'ictor M. Garc\'{\i}a-Su\'arez$^{2}$}
\author{Jaime Ferrer$^1$}

\affiliation{$^1$ Departamento de F\'{\i}sica, Universidad de Oviedo, 33007 Oviedo, Spain}
\affiliation{$^2$ Department of Physics, Lancaster University, Lancaster, LA1 4YB, U. K.}

\date{\today}

\begin{abstract}
We analyze the stability and magnetic properties of infinite zigzag atomic chains of a large number of late 
third, fourth and fifth-row transition metal atoms, as well as of the Group IV elements Si, Ge, Sn and Pb. 
We find that zigzag chains of third- and fourth-row elements are not stable, while those
made of Si, Ge, Sn, Pb, W, Os, Ir, Pt and Au are. These results correlate well with known data in Mechanically
Controllable Break Junction experiemnts (MCBJE). We therefore conjecture that the stability of an infinite chain
is at least a necessary condition for the formation of a finite sized one in MCBJE. We therefore predict that
Sn and Os, and possibly W and Pb chains may be found in those experiments. We also
find that the bonds in Hg chains are extremely soft. We finally show that the magnetic moments and
anisotropies of Ir and Pt chains show a non-trivial behavior.
\end{abstract}


\maketitle

The discovery of free-standing atomic chains (AC) of gold atoms in 1998 \cite{Ohn98,Yan98} has spurred an 
intense experimental and theoretical activity along this decade. Pressing subjects related to their 
structural properties, like the possible geometry of the necks formed at the atomic constriction, the 
actual length of the chains and more importantly the search for other elements that could also form 
AC have been intensively discussed\cite{San99,Bah01}. Despite initial reports on
the formation of AC of several 3d- and 4d-row elements\cite{Uga02,Uga03,Uga04}, unequivocal proof of their 
existence has only been achieved for Au, Pt and Ir\cite{Smi03,Smi03b}. Furthermore, third and fourth-row elements 
like Ni, Co, Rh, Pd and Ag may only form AC upon addition of O$_2$, CO and other gas molecules to the chamber
\cite{Smi03,Smi03b,Thi06}. 
Gold chains have also been shown to display several fascinating transport phenomena, that include conductance 
quantization\cite{Ohn98,Fer88} and oscillations due to parity effects\cite{Lan97}. Parity oscillations
have been proven to happen in Pt and Ir chains also\cite{Smi03}. 

A large number of calculations of AC of either infinite or finite length of a variety of
alkaline or transition metal atoms have been performed to elucidate these fascinating structural and electronic
properties. Unfortunately many of them have restricted either the geometry 
of the chains to linear configurations, or the spin degrees of freedom to the paramagnetic state.
We note that the theoretical ground state of finite- or infinite-sized chains is a zigzag configuration
\cite{San99,Gar05,Jel06,Sen05}. Furthermore, the ground state is frequently magnetic\cite{Del03,Fer05} and
may present sizable magnetic anisotropies\cite{Dor98,Mok06} that could affect the structural,
electronic and transport properties.

We present in this article a thorough study of the structural, electronic and magnetic properties of
infinite AC of the following atoms of the 3d, 4d and 5d series: Co, Ni, Cu, Rh, Pd, Ag, W, Os, Ir, Pt,
Au and Hg, as well as of the Group IV column: Si, Ge, Sn and Pb. This makes a total of 16 elements, that 
constitutes a sizable chunk of the Periodic Table. 
We have used the ab initio LCAO norm-conserving pseudopotential code SIESTA\cite{Koh65,Tro91,SIESTA}, but we have checked that
the plane-wave ultrasoft pseudopotential code Quantum ESPRESSO\cite{PWscf} produces identical results for Si and 
Pb chains. We have used the LDA approximation \cite{Per81} for all elements except for Co and Ni, where 
GGA\cite{Per96} provides a better description of their bulk phase. To ensure that our results are robust
against the choice of exchange and correlation functional, we have also checked that GGA produces
identical results for Si, Sn and Pb chains. We have included non-linear core corrections for most,
but not all, elements to account for the effect of semicore states in the valence. We have used in all 
cases a double zeta polarized basis set with long radii. We have checked that the lattice constant, bulk modulus and 
electronic band structure of the corresponding bulk materials were accurately described. We have repeated 
the calculations of the structural properties of Ni and Pd chains using a triple zeta double polarized basis, and 
have found that the results had converged numerically. Likewise, we have repeated our calculations for Co chains 
with two different pseudopotentials with almost identical results. All our calculations are fully 
non-collinear in the spin degrees of freedom, and include the spin-orbit interaction (SOI)\cite{Fer06}. 
We have additionally performed spin-unpolarized and non-collinear spin-polarized  calculations without SOI for almost 
all the atoms to assess the importance of magnetism and of magnetic anisotropies in the stability and 
geometry of the AC. 

We have chosen a coordinate system where the chains are aligned with the z-axis. To allow for zigzag 
arrangements, we have taken unit cells that enclose two atoms, as shown in Fig. 1. These unit cells are 
parallelograms with a section of 20 x 20 \AA$^2$, and a height that we vary to elongate the chains. 
We have checked in several cases that this section leaves enough vacuum space at the sides to have 
converged results. The position of the two atoms lie initially in arbitrary positions in the xz-plane, 
but we have subsequently optimized the geometry through a conjugate-gradient algorithm that minimizes the 
forces\cite{SIESTA}. We have placed no restriction in the 
position of either of the two atoms, and have set a tolerance in the forces of 0.01 eV/\AA. 
These facts have allowed us to classify the elongation of the zigzag chains according to the projection of
the inter-atomic distance between two atoms along the z-axis, that we call $d_z$, see Fig. 1. Notice that the 
length of the unit cells along the z-axis is {\it twice} $d_z$. We have used a reciprocal meshes of 50-100 k-points and a 
thermal smearing of 50 Kelvin, that has eventually been decreased to 0.1 K to check the dependence of the MAE with the 
thermal smearing. The real-space mesh used to compute the Hamiltonian matrix-elements had from 250 x 250 x 30-90, to 
400 x 400 x 50-120 points in the unit cell, depending on whether LDA or GGA was used, and on the elongation of the chain.

We have found that linear chains are always unstable against lateral displacements of the atoms, that produce zigzag
arrangements. In contrast, these zigzag chains show one or two energy minima as a 
function of $d_z$.  We have found that the absolute minimum corresponds to atomic arrangements where the bonds make 
angles with the z-axis of about 60 degrees, as illustrated in Fig. 1 (a). Further, the 5d elements W, Os, Ir, Pt and Au, 
as well as Si, Ge, Sn and Pb make chains that have the second minimum, while the 3d and 4d elements Ni, Co, Cu, Rh, 
Pd and Ag do not display it. For this second minimum, the angles subtended between the bonds and the z-axis usually are of 
about 30 degrees as shown in Fig. 1 (b). We regard this atomic configuration as a true zigzag atomic chain.
since every atom is chemically bonded only to its two neighboring atoms. On the contrary, we consider that
the atomic configuration in Fig. 1 (a) corresponds better to a two-atom nanowire, since each atom is bonded to more than
two atoms. Fig. 1 (c)  sketches a finite-sized chain similar to the platinum AC that we have simulated recently\cite{Gar05}, 
which describe very well the conductance data of platinum MCBJE. These finite-sized
chains also displayed the two-minimum structure, albeit the zigzag minimum had a reduced depth as compared with the
infinite zigzag platinum chain. Notice that a local minimum indicates a sort of mechanical stability even for a 
nanostructure like
a finite-sized chain. Hence, we expect that only those elements that show it should produce AC in MCBJE.
Further, we found in those finite sized chains that the angles made with the z-axis were 
reduced to about 10-25 degrees, due to the combined effect of the geometric constraints and the strain felt at the
neck. We therefore expect that finite-sized two-atom nanowires can not be realized experimentally due to
the geometry and atom dynamics at the neck. In contrast, the zigzag arrangements of infinite chains have a clear parallel
in actual MCBJE. 
We are therefore led to conjecture that the existence of the zigzag minimum in an infinite chain is at least a 
necessary condition for its appearance in its corresponding finite-sized ACs in MCBJE. 

We note that our predictions correlate very well with the most accurate experimental studies on 
the formation of atomic chains in MCBJE\cite{Smi03,Smi03b}, where out of the nine studied elements Ni, Co, Cu, Rh, Pd, Ag, Ir, 
Pt and Au, only these last three were found to make them. We therefore believe that in addition to Ir, Pt and Au, possibly Os,
and Sn and perhaps Pb and W have the chance to form atomic chains in MCBJE, while it is hard to make any specific 
prediction for Hg. We also 
note that spin-unpolarized or even spin-polarized calculations with no SOI overestimate the stability of the chains for some heavy elements, hence a fully spin-polarized is often required to calculate this stability correctly. 
We confirm that Pt and Ir chains are magnetic, while the magnetism of Au AC appears only for very elongated chains and
is very weak\cite{Del03,Fer05}. In addition, we find that Os and Hg chains display magnetism at different 
elongations, while Pb and Sn AC only magnetizeat very long $d_z$. Furthermore, only Pt and Ir chains are magnetic at their 
equilibrium configuration. We finally study the magnetic anisotropy energy (MAE) of Au, 
Pt and Ir AC. We find that the easy axes of Ir and Pt lie perpendicular and parallel to the chains, respectively, while
Au has no appreciable MAE. 

\begin{figure}
\includegraphics[width=0.9\columnwidth]{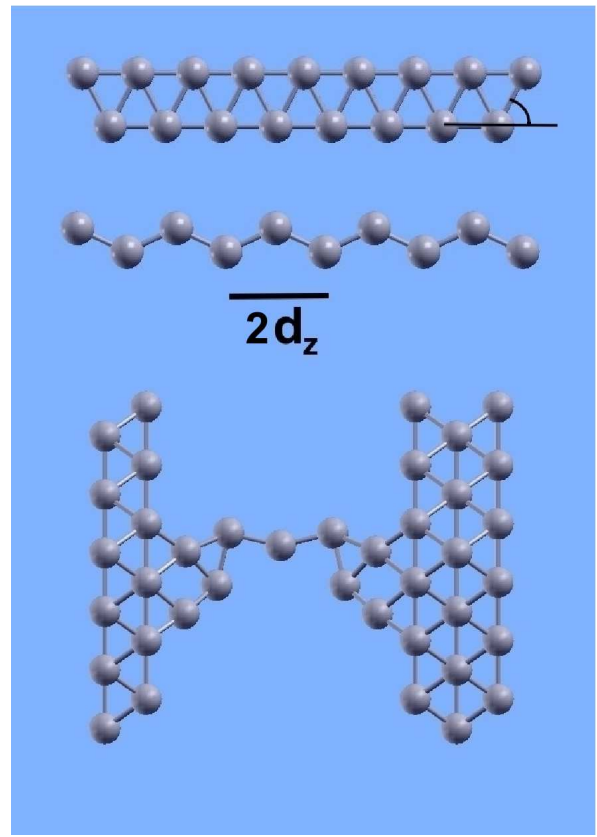}
\caption{(a) Infinite two-atom nanowire, making angles of about 60º with the wire axis. (b) Infinite zigzag chain, that
makes angles of about 30 degrees with the chain axis. $d_z$ is the projection of the interatomic length into the chain 
axis, e.g.: half of the unit cell length along the axis. 
(c) Initial stages of a five-atoms finite-sized 
platinum chain.}
\end{figure}

We present first our study of the Group IV elements Si, Ge, Sn and Pb, that we use partly to test our 
methodology, since the first three have been studied already by Senger and coworkers\cite{Sen05}.
We have simulated zigzag chains with SIESTA using both LDA and GGA, and allowing for spin polarization; 
we have repeated the calculations for Sn and Pb taking into account the SOI. Aditionally, we have also 
simulated paramagnetic linear and zigzag chains with Quantum ESPRESSO using GGA\cite{PWscf}.
Our results are summarized in Fig. 2, where the energy per atom, the tension defined as
\begin{equation}
T=\mathrm{d}E/\mathrm{d}d_z
\end{equation}
and the angles subtended by the bonds and the z-axis are plotted as a function of $d_z$. Note that the 
energy curves at the top panels have been shifted rigidly to ease visual comparison of the data. Further,
we present only a suitable selection of the collected data  that illustrate the following discussion, to avoid
confusing the reader with too many curves.
We have found that SIESTA and Quantum ESPRESSO provide identical results for the properties of these chains. Furthermore, 
these results are essentially the same regardless of the approximation used for exchange and correlation (LDA 
or GGA). We note that inclusion of the SOI does not modify the energy curves of Sn. In contrast, inclusion
of the SOI clearly modifies the depth of the zigzag minimum for Pb, actually reducing it. In other words, a
spin unpolarized (or spin polarized with no SOI) calculation would overestimate the stability of zigzag chains.
We have noticed that a plot of the chain tension\cite{Rub96,Tos01} is a better guide to estimate such stability, see 
the middle panels in Fig. 2. These graphs clearly demonstrate that the two codes and whichever
approximation taken for exchange and correlation provides essentially identical results for the structural properties
of zigzag chains. We note that $T$ is negative for small $d_z$ but eventually becomes positive and that 
the position of the absolute minimum is given
by the first crossing through zero. Further, the existence of the zigzag minimum requires that $T$ becomes again negative,
the position of the minimum corresponding with the third zero crossing. Notice also that the larger the depth of the 
local energy minimum, the more negative $T$ becomes. We therefore
find that the stability of zigzag chains is largest for Si and gradually decreases to a small but finite value
in Pb (which is further reduced by Spin-Orbit effects). We also note that the energy curves for zigzag chains 
have a kink at the point where they merge with linear chain curves. This kink translates to a step in the $T$ curves. 
We have found that the bond angles fall precipitously to zero at such points and the chains
become linear, as shown in the bottom panels of Fig. 2.  Notice also that the two energy minima parabola 
(nanowire and zigzag) do not
match smoothly. In other words, the maximum that exists in between is too sharp, indicating that the transition
between the two structures is rather abrupt. This is also manifested in the kink shown at
the minimum of the tension, and in the sudden drop shown for the angles at the same distance. Notice finally that
the angles that correspond to the zigzag minimum are of about 25-30 degrees.
Following our conjecture, we expect that Si, Ge and Sn could form atomic
chains in MCBJE, while the case of Pb is more uncertain, since its tension becomes negative indeed, but with a
small value. Since Si and Ge are semiconductors, it seems that only Sn (and perhaps Pb) chains could be 
detected in those experiments. 

\begin{figure}
\includegraphics[width=0.95\columnwidth]{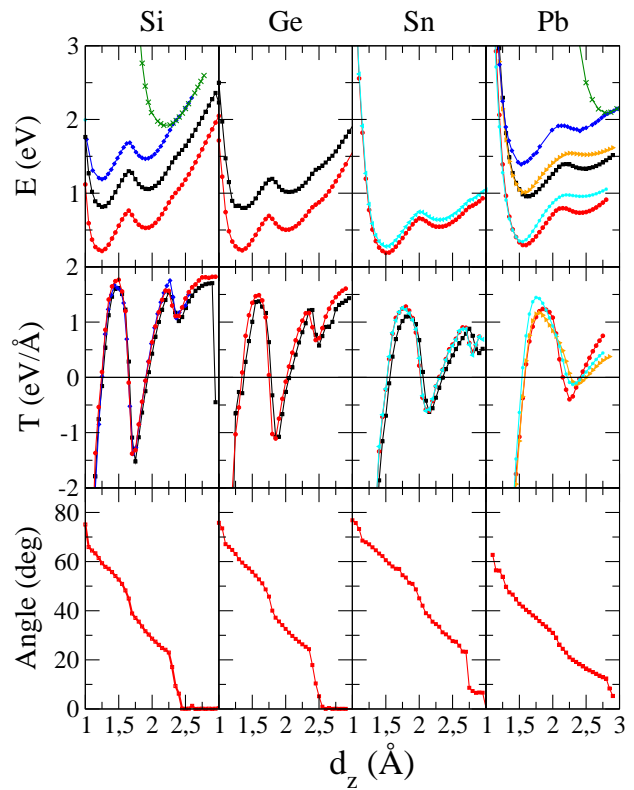}
\caption{Top panels: energy per atom of Si, Ge, Sn and Pb zigzag and linear chains. 
Curves have been rigidly shifted to aid comparison among them. Spin polarized SIESTA LDA curves for the 
four elements are plotted at the bottom of the graphs; Sn and Pb show also spin polarized curves with
SOI included, that are almost superimposed to the former. Si, Ge and Pb show at medium height spin polarized
SIESTA GGA curves; Pb also shows the curve with SOI included. Si and Pb show at the highest height the spin
unpolarized PWscf GGA results for zigzag and linear chains. 
Middle panels: tension of zigzag chains of Si, Ge, Sn and Pb.  For Si, Ge and Sn all curves basically collapse in a single
line. Pb only shows the curves for spin polarized SIESTA LDA with and without SOI, and SIESTA GGA with SOI, where it
is demonstrated that SOI reduces the stability.
Bottom panels: angles
subtended between the bonds and the chain axis for Si, Ge, Sn and Pb zigzag chains.
}
\end{figure}

We discuss now the stability and geometry of Ir, Pt and Au chains. 
We have proven that the results of our simulations are robust against the choice of code and exchange
and correlation approximation. Hence, the calculations described now have been performed with SIESTA in the LDA 
approximation, which provides a slightly more accurate description of the lattice constant in the bulk phase of these
materials. We have performed simulations of linear and zigzag chains for Ir and Pt. We have looked for paramagnetic 
as well as for spin polarized solutions, these last ones with and without SOI. Our results are summarized in Fig. 3.
The top panels show how the energy per atom behaves as a function of $d_z$. These curves show again the two-well 
structure and the kink and the merging point between linear and zigzag chains. Clearly, spin unpolarized calculations
overestimate the stability of Ir and Pt chains, as better seen in the panels for the tension. The shift observed between
the energy curves with and without spin orbit also indicate the size of Spin-Orbit effects  for Ir and Pt chains.
In contrast, these effects are much smaller for gold, as will be discussed later on in this article. 
In any case, the tension of these three materials attain negative values of about 1 eV/\AA, similar to Sn. The angles made 
by Ir are a 
bit large, of about 45 degrees, while for Pt and Au, these angles are much lower, about 25 and 20 degrees, respectively.
Notice also that the sudden drop in angles seen at the merging point of the two energy minima parabola in Group IV elements
is smoother (but noticeable) now.
According to our conjecture, these three elements should make atomic chains, as they indeed do. 

\begin{figure}
\includegraphics[width=0.95\columnwidth]{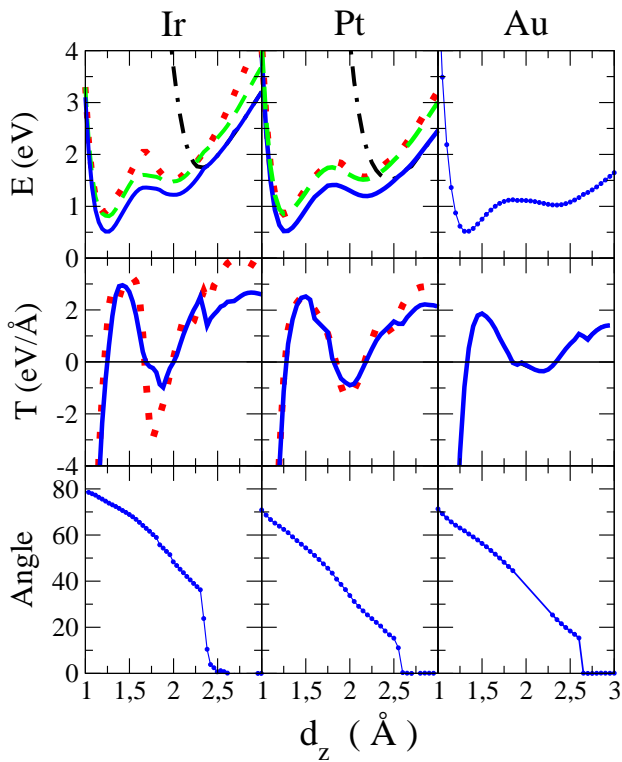}
\caption{
Top panels: energy per atom of Ir, Pt and Au zigzag chains, calculated with SIESTA. These calculations
are spin polarized and take into account the Spin Orbit interaction. For Ir and Pt we also show the
corresponding results for linear chains. We also show for Ir and Pt the results for spin unpolarized and
spin polarized without SOI zigzag chains.
Middle panels: tension as a function of $d_z$.  Ir and Pt show results for spin unpolarized and spin polarized
with SOI, to stress how this the first overestimate the stability of the chains.
Bottom panels: angle subtended by the bonds and the z-axis in Ir, Pt and Au zigzag chains.
Solid and dashed lines indicate spin polarized calculations with and without SOI, respectively; dotted lines represent
spin unpolarized simulations; dash-dotted lines indicate calculations for linear chains (with SOI).
}
\end{figure}

We now undertake a survey of the tension in zigzag AC of the 3d-row elements Co, Ni and Cu, and 4d-row
Rh, Pd and Ag. Detailed MCBJ experiments\cite{Smi03,Smi03b} have shown that these elements do not form finite 
length chains. Our results from
spin-polarized plus SOI simulations are shown in Fig. 4. We find that $T$ never becomes negative for
Co, Ni and Cu, nor for Rh, Pd and Ag, in contrast with Ir, Pt and Au. Among the third-row elements, the
tension of Co is not only positive, but also pretty large always, while $T$ for Cu, and specially for
Ni attains small positive values in a small window of elongations about 1.75 \AA. Among the 4d-row elements, 
the tension of Pd is also large for all elongations. We wish to stress that we have repeated the calculations 
for Ni, Co and Pd with different basis sets or pseudopotentials to confirm their behavior. 
The case of silver is interesting since, even though $T$ is never negative, its magnitude is small 
for a longer range of elongations. These results correlate very well with the stability of AC found in actual 
MCBJ experiment. These results reinforce our belief that the existence of the zigzag minimum is at the very least a necessary
condition for the formation of atomic chains in MCBJE. Even though it is perfectly clear that our simulations 
are no substitute for those of an actual atomic
constriction\cite{Gar05,Fer05, Jel05}, we propose to use the tension of an infinite chain as a quick but
reliable criterion to predict the actual formation of finite-length chains in MCBJ experiments.

\begin{figure}
\includegraphics[width=0.9\columnwidth]{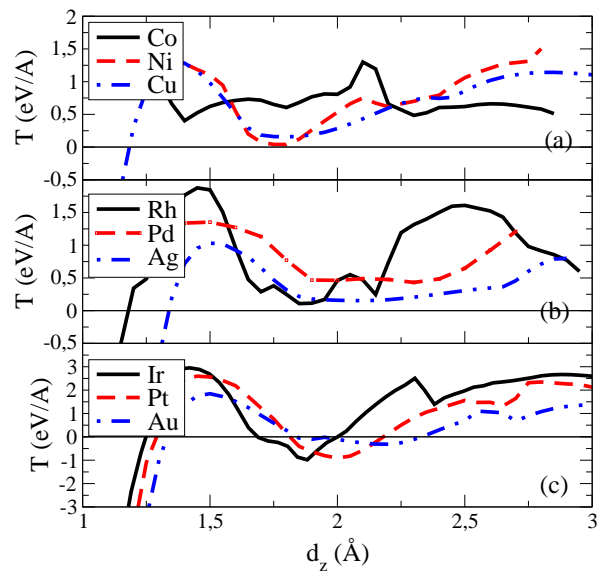}
\caption{
(a) Tension in Co (solid), Ni (dashed) and Cu (dash-dotted) zigzag chains as a function of $d_z$. 
(b) Same for Rh (solid), Pd (dashed) and Ag (dash-dotted). 
(c) Same for Ir (solid), Pt (dashed) and Au (dash-dotted).
}
\end{figure}

The enhanced stability of 5d-row transition metal AC has been attributed to relativistic effects, that
shift down in energy the s band, while moving up the top of the d bands\cite{Tak89}. Clearly, this is not
a universal effect, since exactly the contrary happens for Group IV elements. It is likely in any case that not only 
Ir, Pt and Au but also other fifth-row atoms may form chains in MCBJ experiments. 
We present our results now for the tension of chains of the fifth-row elements W, Os and Hg, that we
show in Fig. 5. We have found that the tension of W and Os becomes even more negative than that of Ir, Pt and Au.
But we find that the angles made by W at the zigzag minimum are large, of about 45 degrees. The angles for osmium are
much lower, of about 30 degrees, so Os AC can be truly regarded as zigzag chains, while W seems closer to nanowires. 
The case of Hg is very interesting, since it makes extremely soft bonds. Our survey of the 5d row uncovers also that 
the propensity to form chains is a monotonically decreasing function of the atomic number within the row. It is 
also clearly related to the hardness of the materials but maybe in a non-trivial way, since Os seems to be harder 
that W\cite{Kle06}.  These results seem to indicate that Os and perhaps W could most possibly form true chains in MCBJE, while it is 
not clear at all what could happen experimentally with Hg. Atomic chains of Hg could have very interesting elastic 
and plastic properties, in case they would be shown to exist.

\begin{figure}
\includegraphics[width=0.9\columnwidth]{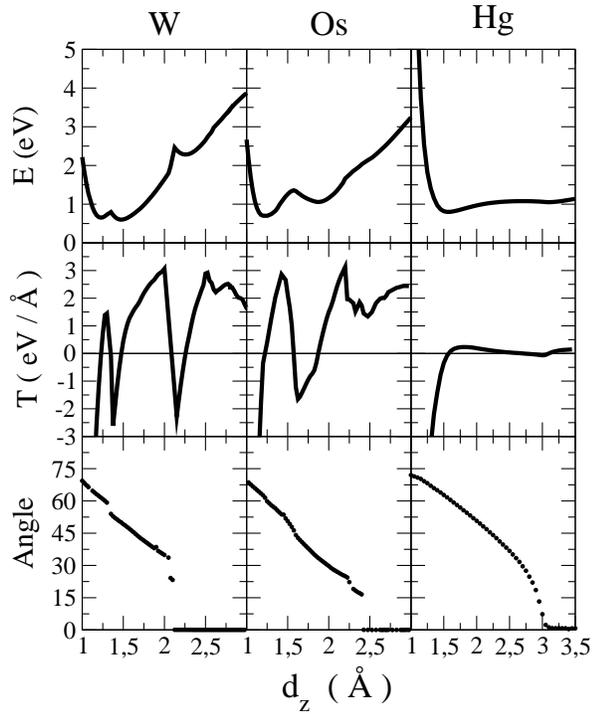}
\caption{
Top panels: energy per atom of zigzag chains of W, Os and Hg as a function of $d_z$.  Spin polarized calculations with SOI
included have been performed.
Middle panels: tension in zigzag W, Os and Hg chains as a function of $d_z$. 
Bottom panels: angle subtended by the bonds and the z-axis in W, Os and Hg zigzag chains.
}
\end{figure}

We turn now the attention to the magnetic properties of the stable zigzag Ir, Pt and Au chains. Fig. 6
shows their magnetic moments and magnetic anisotropies as a function of $d_z$.  Our calculated moments 
for linear chains agree well with the results of Delin and Tosatti\cite{Del03}. The magnetism of zigzag Ir chains
is interesting, since it shows High-Spin to Low-Spin to High-Spin transitions upon stretching in the
range of elongations that
correspond to the zigzag minimum. The magnetism of Au chains, and its accompanying MAE is negligible.
Magnetic anisotropies of (theoretically unstable) linear chains of 4d transition metal atoms have been 
calculated recently by Mokrousov and coworkers\cite{Mok06}, who found that the magnetic moment aligns 
parallel to the axis of the chains for some elements, and perpendicular to it for others. Notice that an 
ambiguity arises in the last case for zigzag chains since the x- and y- axes are now inequivalent. 
Indeed, we find that the magnetization vector of (also unstable) zigzag Pd chains lies along the x-axis, e. g.: in the
plane of the zigzag. Our results for the MAE of the theoretically stable Ir, Pt and Au zigzag chains are shown
in Fig. 6. We define E$_i$ ($i\,=\,x,\,y,\,z$) as the energy of a zigzag chain where all atomic spins are aligned 
along the $i$-axis. Performing all possible substractions allows to define the MAE.
We note that the equilibrium $d_z$ of these chains are, respectively, 1.8, 2.0 and 2.2 \AA.
Consequently the easy axis of Ir is aligned with the x-axis, with MAEs of 20 and 15 meV with respect to the z-
and y-axes. On the contrary, the easy axis of Pt is aligned with the chain axis, and the MAE is somewhat smaller,
of about 12 meV.

\begin{figure}
\includegraphics[width=0.9\columnwidth]{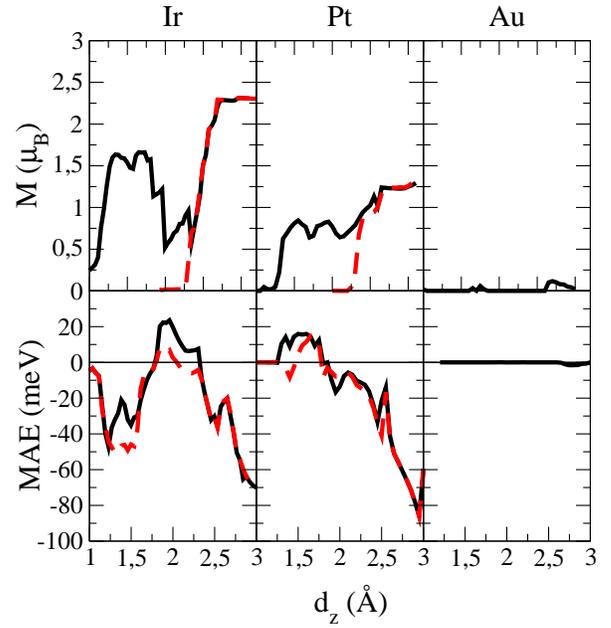}
\caption{
Top panels: magnetic moments per atom of zigzag (solid) and linear (dashed) chains of Ir, Pt and Au.
Bottom panels: 
MAE of Ir, Pt and Au zigzag chains as a function of $d_z$. 
The zigzag chains lie in the xz plane. Solid lines indicate the in-plane anisotropy E$_z$-E$_x$; 
dashed lines, the out-of-plane anisotropy E$_z$-E$_y$.
}
\end{figure}

To summarize, we have computed the structural, electronic and magnetic properties of
infinite zigzag atomic chains made of Group IV elements and a variety of late transition metal elements.
We have found that the energy versus elongation curves have an absolute minimum that corresponds to two-atom
nanowires. In addition, some elements show a second local minimum for longer elongations which corresponds
to true zigzag chains. We conjecture that the existence of this minimum is at the very least a necessary condition,
and possibly also a strong signature, for the formation of atomic chains in MCBJE. We indeed find that Ni, Co, Cu, Rh, 
Pd and Ag do not have the 
minimum, while Ir, Pt and Au do have it, in very good agreement with MCBJE\cite{Smi03,Smi03b}.
We confirm that Si, Ge and Sn also display the minimum\cite{Sen05} as do Pb, W and Os. Since Si and
Ge are irrelevant for MCBJE, we predict that Sn and Os, and perhaps W and Pb may also form those chains.
Sn and Pb are specially interesting since the bulk materials become superconducting at low temperatures.
These MCBJE could therefore become Josephson junctions were the weak link is an atomic chain of tunable length.
The case of Hg is also interesting, since the infinite chains make very soft bonds, but it is not clear whether 
they could be realized experimentally. We find that the magnetic anisotropy of gold chains is very weak, 
while Pt and Ir have a stronger MAE, their easy axes lying parallel and perpedicular to the axis of the
chain, respectively.
 
\section*{Acknowledgments}
It is our pleasure to acknowledge useful conversations with P. Ordej\'on, J. M. Soler,
D. S\'anchez-Portal, J. J. Palacios, A. Vega, S. Sanvito, 
M. A. Oliveira, A. Postnikov, C. J. Lambert, J. van Ruitenbeek and specially C. Untiedt, A. Halbritter
and Sz. Csonka. We also acknowledge financial support from the EU network MRTN-CT-2003-504574
RTNNANO and the Spanish project MEC BFM2003-03156. LFS is funded by
FICYT grant BP04-087 while VMGS thanks the European Union for a Marie Curie grant.
A. Halbritter and Sz. Csonka reminded to us that some of these materials are superconductors.

\end{document}